\documentclass{elsart3p}

\usepackage{graphics}
\usepackage{graphicx}
\usepackage{epsfig}

\usepackage{amsmath,amssymb}

\usepackage{units}
\usepackage{multicol}
\usepackage{multirow}
\usepackage{subfigure}
\usepackage{textcomp}
\usepackage{xspace}

\makeatletter
\def\cleareads{%
  \csname email@text\endcsname={}%
  \csname url@text\endcsname={}%
  \csname corauth@text\endcsname={}%
  \global\chardef\has@ead@email=0\relax
  \global\chardef\has@ead@url=0\relax
  }
\makeatother

\def\gcm2{g/cm$^2$\xspace}

\newtoks\pnam
\def\Label#1{\label{\the\pnam#1}}
\def\Ref#1{\ref{\the\pnam#1}}
\def\Cite#1{\cite{\the\pnam#1}}
\def\Bibitem#1{\bibitem{\the\pnam#1}}

\def\theseproc#1{Proc. of the 5th Fluorescence Workshop, Nucl. Instr. \&
                 Meth. A (2008) in press.}

\begin{document}

\begin{frontmatter}


\title{Air Fluorescence Relevant for Cosmic-Ray Detection --- Review of
       Pioneering Measurements}


\author[m]{Fernando Arqueros},
\author[n]{J\"org R. H\"orandel},
\author[k]{Bianca Keilhauer}

\address[m]{Facultad de Ciencias F\'{\i}sicas,
Universidad Complutense de Madrid, E-28040 Madrid, Spain}
\address[n]{Radboud Universiteit Nijmegen, Department of Astrophysics,
             P.O. Box 9010, 6500 GL Nijmegen, The Netherlands}
\address[k]{Universit\"at Karlsruhe, Institut f\"ur Experimentelle
            Kernphysik, Postfach 3640, 76021 Karlsruhe, Germany}

\begin{abstract}
 Cosmic rays with energies exceeding $10^{17}$~eV are frequently registered by
measurements of the fluorescence light emitted by extensive air showers.  The
main uncertainty for the absolute energy scale of the measured air showers is
coming from the fluorescence light yield of electrons in air.  The fluorescence
light yield has been studied in laboratory experiments.  Pioneering
measurements between 1954 and 2000 are reviewed.
\end{abstract}

\journal{Nuclear Instruments and Methods A}
\end{frontmatter}

\section{Introduction}
\Label{intro}

Cosmic rays with energies exceeding $10^{17}$~eV are frequently observed by
measurements of the fluorescence light emitted in air showers. The latter are
induced by high energy cosmic rays interacting in the Earth's atmosphere
initiating a cascade of secondary particles. Relativistic electrons (and positrons) are
the most numerous charged particles in air showers. On their way through
the atmosphere they excite nitrogen molecules. The nitrogen molecules release
their excitation energy partly through isotropic emission of fluorescence
light.  The fluorescence light is measured with imaging telescopes in air
shower experiments, allowing for a three-dimensional reconstruction of the
shower profile in the atmosphere. The main uncertainty of the absolute energy
scale of the measured showers is coming from the fluorescence light yield
of electrons in air. 

The bulk of the fluorescence light is induced by electrons with MeV energies.
Thus, the emission mechanism can be studied in laboratory experiments using
particles, mainly electrons, from radioactive sources or particle
accelerators. Particles are injected into volumes of nitrogen and air under well
controlled circumstances and the fluorescence light yield is measured as a function of
various parameters, like electron energy, gas pressure, temperature, and
humidity.

Recent developments have been discussed during the 5th Fluorescence Workshop in
El Escorial, Spain from September 16th to 20th, 2007.  The results are
summarized in an accompanying article \Cite{summary}.  The present article
summarizes measurements of the fluorescence light yield conducted between 1954
and 2000.  The objective is to compile information relevant for air shower
observations from the (sometimes hardly accessible) historical papers.  Their
implications on the contemporary understanding of the subject are discussed in
\Cite{summary}.

\Label{sec:fl_history}

\section{A.~E.~Gr\"un, E.~Schopper, (1954)}
\Label{sec:grun_schopper}

Very early investigations of fluorescence emission in gases induced by rapid particles
were performed by Gr\"un and Schopper~\Cite{grun_schopper1954} in 1954. In the
experimental setup, described in the paper, $\alpha$-particles from a radioactive
Po-source were used to excite the gas. The measurements were done mainly
in nitrogen gas and nitrogen argon mixtures, but also in nitrogen gas with different
admixtures like Xe, O$_2$, etc. The fluorescence yield $\Phi$\footnote{In the
nomenclature of the summary article \Cite{summary},
the $\Phi(N,T,c,v)$ from Gr\"un and Schopper corresponds to
$\Phi_\lambda$.} was defined as
\begin{equation}
\Phi(N,T,c,v) = \frac{\phi}{\epsilon},
\end{equation}  
where $\phi$ is the radiation power in W/cm$^3$ and $\epsilon$ is the de-acceleration
power, also in W/cm$^3$, of the primary particle stream in the gas. The fluorescence yield
depends on $N$, the particle number density of the gas, $T$ is the temperature of the gas,
$c$ is 
the concentration of the admixture to the nitrogen, and $v$ is the velocity
of the exciting particles as it enters the gaseous volume. The radiation power was
measured photo-electrically with a photomultiplier 1 $P$ 28. The experimental setup was
chosen carefully, since the measured stream of photons yields only the fluorescence yield
under the following conditions, as stressed in the publication~\Cite{grun_schopper1954}:
\begin{enumerate}
\item The gas volume has to be independent of the variable which is varied during
the measurement. Then the proportionality between the measured photo-stream and the
radiation power $\phi$ is fulfilled.
\item The dependence of the decelerations power $\epsilon$ on the varied variable has
to be known. The derived ratio $\phi/\epsilon$ is proportional to the outer fluorescence
yield.
\item The emitted radiation must not be absorbed by the gas itself. This ensures that the
outer yield is equal the inner fluorescence yield.
\item The spectral combination of the radiation must be constant because of the selective
sensitivity of the photo cathode.
\end{enumerate}
Additionally, the setup was such that no $\alpha$-particle hits any wall within the
measurement volume. To account for deceleration of the exciting particles before and in
the observed volume under high pressure conditions, a pressure-dependent correction factor
was introduced.

Firstly, the pressure-dependence of the fluorescence yield $\Phi(p)$ was measured for
constant temperature $T$. The fluorescence yield was written in dependence on
pressure $p$\footnote{In the 
nomenclature of the summary article \Cite{summary},
the $p$ from Gr\"un and Schopper corresponds to
$P$ and $p^\prime$ to $P^\prime_{v^\prime}$.} as
\begin{equation}
\Phi(p)=\frac{\Phi_0}{1+p/p^\prime} \\
{\rm{with}}~p^\prime=\frac{p_0\gamma_0}{\delta(T)N_0}
= p_0\frac{\tau_L}{\tau_0}, 
\end{equation} 
with $\tau_L=1/\delta N_0$ as the mean lifetime of the excited molecule with respect to
quenching. $\Phi_0$ is the yield for $N \rightarrow 0$ meaning maximal yield. 
$\gamma_0$ is defined as $1/\tau_0$, where $\tau_0$ is the mean lifetime without quenching.
$\delta$
means the ``quenching velocity'' and can be given as 
\begin{equation}
\delta = \delta (T) = u(T)\cdot \sigma_L(T).
\end{equation}
$u(T)$ is the mean relative velocity of the gas molecules and $\sigma_L(T)$ is the
temperature-dependent quenching cross section. In pure nitrogen, the fit to the data
points led to a value of 
$p^\prime_{{\rm{N}}_2}$ = 35~Torr~\Cite{grun_schopper1954}. The authors concluded a ratio
between $\tau_L$ and $\tau_0$ of 0.05 at 760~Torr and 20$^\circ$C. The estimated mean
lifetime $\tau_L$ of the C$^3\Pi_{\rm{u}}$-state of N$_2$-molecules with respect to
quenching was about 5$\cdot10^{-10}$~sec with $\tau_0 \approx 10^{-8}$~sec. It was pointed
out that the quenching cross section $\sigma_L$ is smaller by a factor of 1/2.5 compared
with the gas-kinetic cross section of N$_2$-molecules.

Secondly, the temperature dependence of the fluorescence yield was discussed. It was
stated that this dependence is caused by the temperature dependence of the quenching
process. Measured data were plotted for the function $\delta (T)$. The authors
concluded that the quenching cross section $\sigma_L$ decreases with increasing
temperature apart from a $\sqrt T$-increase of the mean relative velocity $u$ of
molecules with increasing temperature.

Thirdly, the role of secondary electrons and the concentration-dependent light yield of
Ar-N$_2$-mixtures was analyzed~\Cite{grun_schopper1954}. The spectroscopic investigation 
started with pure argon
and N$_2$ was added successively. In pure Ar, the radiation power stems from Ar spark
lines induced by ionization. In pure nitrogen, the emission comes mainly from the second
positive system and smaller contributions from N$_2^+$. It was found that already at small
concentrations of N$_2$, the light yield is significantly increased. The authors explained
it by an energy transfer from Ar to N$_2$. The deceleration power of the incoming particles
is absorbed by both gases and part of the energy absorbed by argon is transferred to
nitrogen 
which then radiates fluorescence light. The idea of energy transfer is supported by the
fact that the C$^3\Pi_u$-state of N$_2$ cannot be excited directly by primary
particles. The obtained concentration-dependent light yield curve is the result of the
concentration dependence of the energy transfer and of concentration-dependent
quenching. The steep increase of the yield curve already at small nitrogen concentrations
was explained by the authors with large cross sections for the process of energy
transfer. With increasing nitrogen concentration, the energy transfer saturates and
decreases afterwards because of the decreasing Ar-concentration. Additionally, the
quenching due to argon in the nitrogen gas superimposes. In conclusion, it was stated that
the 
energy transfer occurs mainly due to secondary electrons from argon.

\section{A.~E.~Gr\"un, (1958)}
\Label{sec:grun}
Three years later, Gr\"un published measurements of fluorescence yield in
air~\Cite{grun1958}. He used a parallel, narrow beam of electrons with energies between
4.2 and 43.7 keV. The aim of the
investigation was to determine the light yield as a function of the gas density. Therefore,
the function $\Phi(p)$ was measured at constant temperature $T = 20^\circ$C within the
pressure range from 10 to 600~mm~Hg. From a description of the fluorescence mechanism,
Gr\"un deduced the proportionality\footnote{For the nomenclature of the summary article \Cite{summary},
see footnotes of Sec.~\Ref{sec:grun_schopper}.}
\begin{equation}
1/\Phi \propto 1 + p/p^\prime.
\end{equation}
A linear fit to the data points yielded a $p^\prime = 11.5$~mm~Hg for air as an average over
a large spectral range.

In the following years, the aurora phenomena in the Earth's atmosphere came more and
more into focus. Several investigations about light yield, excitation and de-excitation
processes were carried out. These studies were performed in pure nitrogen or air, because
also the aurora phenomena are mainly nitrogen scintillation high up in the
atmosphere. Here, only some of the experiments are mentioned exemplarily whose results are
used later in the discussions about nitrogen fluorescence induced by extensive air
showers. 

\section{B.~Brocklehurst, (1964)}
\Label{sec:brocklehurst}

In 1964, Brocklehurst published quenching parameters for nitrogen-nitrogen and
nitrogen-oxygen collisional de-excitation~\Cite{brocklehurst1964}. To this end, he measured the
light emission in pure nitrogen gas and in nitrogen-oxygen mixtures within a pressure
range of 1 to 750~mm~Hg. The excitation was performed by soft X-rays. The spectrum was
presumably continuous between 5 and 45~keV with a maximum intensity at about 25 to 30~keV.
The emission spectra were observed using a spectrograph and intensity measurements were
done with a 1 $P$ 28 photomultiplier. Main contributions to the light yield were
identified to be of the second positive bands ($C^3\Pi_u \rightarrow B^3\Pi_g$) of N$_2$ 
and of the
first negative bands ($B^2\Sigma_u^+ \rightarrow X^2\Sigma_g^+$) of N$_2^+$.

For studying the kinetics of collisional processes, the pressure was chosen to
be higher than 1~mm~Hg, since above this value the excitation conditions are
independent of pressure~\Cite{brocklehurst1964}. It is mentioned that secondary
electrons cause excitation above 10$^{-2}$~mm~Hg, but are completely absorbed
in the gas for pressure higher than 1~mm~Hg. Furthermore, the excitation is
assumed to be without ion recombination processes. Former measurements are
cited in the publication by Brocklehurst~\Cite{brocklehurst1964} showing that
fields applied for the removal of ions do not have any effect on the light
output. An iterative procedure was applied in the analysis of the quenching
parameters for separating between 1~N and 2~P states of nitrogen. The factor
$k$, which is given in the publication, corresponds to $P^\prime$ in modern
nomenclature.  The necessary fits to the data were done in the low-pressure
range up to 150~mm~Hg for 2~P and up to 40~mm~Hg for 1~N. For higher pressure,
quenching effects make the data points deviate from the linear dependence found
at low pressure. To extract collisional cross sections, the author used the
lifetimes from 
Bennett and Dalby~\Cite{bennett_dalby1959} which are (4.45 $\pm$ 0.6) $\cdot 10^{-8}$~sec
for the 2P system and (6.58 $\pm$ 0.35) $\cdot 10^{-8}$~sec
for the 1N system . The results are listed in
Table~\Ref{tab:brocklehurst}.
\begin{table}[t]
\caption{Quenching parameters given in Brocklehurst,
1964~\Cite{brocklehurst1964}.\Label{tab:brocklehurst}} 
\begin{center}
\begin{tabular}{ccccc}\hline
upper & $v^\prime$ & quencher & $k$ & $\sigma$ \\ 
state & & & ~~~~~~mm$^{-1}$~~~~~~ & \AA{}$^2$ \\ \hline
$C^3\Pi_u$ & 0 & N$_2$ & 0.015 $\pm$ 0.002 & 0.38 $\pm$ 0.09 \\
 & 1 & & 0.041 $\pm$ 0.004 & 1.04 $ \pm$ 0.22 \\
 & 2 & & 0.047 $\pm$ 0.010 & 1.2 $\pm$ 0.35 \\
 & 0, 1 & O$_2$ & 0.45 $\pm$ 0.15 & 12 $\pm$ 5 \\
$B^2\Sigma_u^+$ & 0 & N$_2$ & 0.85 $\pm$ 0.3 & 15 $\pm$ 5 \\
 & 0 & O$_2$ & 0.8 $\pm$ 0.4 & 14 $\pm$ 7 \\ \hline
\end{tabular}
\end{center}
\end{table}

\section{G.~Davidson, R.~O'Neil, (1964)}
\Label{sec:davidson_oneil}

For determining the characteristics of the spectra radiated from certain gases, Davidson
and O'Neil bombarded nitrogen and air with 50~keV
electrons~\Cite{davidson_oneil1964}. The main results have been the efficiencies for the
 conversion of
electron energy to optical radiation for about 100 transitions in air and nitrogen between
320 and 1080~nm. The pressure for this investigation was fixed at 600~Torr. The radiation
was spectrally analyzed by a monochromator with either a RCA 6199 photomultiplier with
S-11 response or a liquid nitrogen cooled RCA 7102 photomultiplier with S-1 response. The
system was calibrated absolutely using a standard tungsten ribbon filament lamp and a
1000$^\circ$C blackbody radiation. The absolute fluorescent efficiency for a given spectral
component was defined as the ratio of power radiated by the spectral feature to the
incident electron beam power. The estimated error for the efficiency was determined to be
$\pm$ 
15\% due to the absolute radiance calibration and the beam current measurement. The
authors listed the integrated intensities of the observed spectra in terms of absolute
fluorescent efficiencies in a table~\Cite{davidson_oneil1964}.
The given wavelengths are band-head wavelengths after Wallace~\Cite{wallace1962}. Since
the data of the publication \Cite{davidson_oneil1964} were updated in the full report
about the work of Davidson and O'Neil \Cite{oneil_davidson1968}, here no data are reviewed
and the reader is referred to Sec.~\Ref{sec:oneil_davidson}.

\section{A.~N.~Bunner (1967)}
\Label{sec:bunner}

The first study of nitrogen fluorescence emission with respect to cosmic ray detection
were performed by A.~N.~Bunner in the mid 1960s, when Bunner published his thesis
(1964)~\Cite{bunner1964} 
and PhD thesis (1967)~\Cite{bunner1967}. Bunner accurately discussed excitation
and de-excitation processes. It is stressed that for pressure down to values
corresponding 
to atmospheric altitudes of about 60~km a.s.l., the fluorescence spectrum consists almost
entirely of the second positive N$_2$ system (2P) and the first negative N$^+_2$ system
(1N). 

The upper level of 1N can be excited directly
\begin{equation}
{\rm N}_2 + e \rightarrow {\rm N}^{+*}_2 +e +e .
\end{equation}
However, the upper levels of the 2P system cannot be excited directly by high energy
interactions because of forbidden changes in the electronic spin. At lower velocities of
the incoming electrons, the resultant spin change is possible, which is tantamount to
excitation by low energetic secondary electrons. Also cascading from higher levels may
excite the 2P system:
\begin{eqnarray}
{\rm N}_2 + e (\uparrow) &\rightarrow& {\rm N}_2^*(^3\Pi_u) + e(\downarrow) ~~{\rm or} \\
{\rm N}_2^+ + e &\rightarrow& {\rm N}_2^*(^3\Pi_u).
\end{eqnarray}
In the absence of collisional quenching, the absolute intensity of a band is proportional
to $N_{v^\prime} A_{v^\prime v^{\prime\prime}}$, where $N_{v^\prime}$ is the number
of molecules per unit volume in the state $v^\prime$, and
$A_{v^\prime v^{\prime\prime}}$ is the transition probability for a radiative
transition from the level $v^\prime$ to the level $v^{\prime\prime}$, so-called
Einstein coefficients. For an approximate calculation of the distribution of populations
in various excited states, the Frank-Condon principle was used, which means that molecular
electronic transitions occur so rapidly that the internuclear separation stays
constant. Bunner summarized several excitation channels by the following equation:
\begin{equation}
\begin{split}
\frac{{\rm 2P excitations}}{{\rm cm}} &= N\sigma_{{\rm 2P}}^e(E) \\
&+ N \int_0^{E^\prime_{max}} \sigma_{ion}^e(E,E^\prime) S(E^\prime) dE^\prime\\ 
 &+ N \sigma_A^e (E) S(E_A).
\end{split}
\end{equation} 
The first term describes the number of excitations of level $v$ per cm$^3$ and second,
where $N$ is the number of molecules per cm$^3$ and $\sigma^e_{{\rm 2P}}$ gives the
cross section for excitation of a level $v$ by electron incident with energy $E$. The
second term describes the number of excitation from secondary electrons produced by
ionization. The ionization potential of N$_2$ is 15.7~eV, whereas the average energy loss
per ion pair in air is about 34~eV. Thus, the ionization results in a distribution of
secondary electrons, so-called delta rays, having typically several eV energy, sufficient
to excite both the B$^2\Sigma^+_u$ level and the C$^3\Pi_u$ level of
N$_2$~\Cite{bunner1967}. The third term accounts for the excitations from Auger
electrons. Since a high energy electron has about equal probability of interacting with
any atomic electron, a certain number of ionizations will liberate K-electrons and lead to
the emission of Auger electrons, which on their part can excite nitrogen.

\begin{table*}[t]\centering
\caption{Deactivation constants for nitrogen and air given in
Bunner, 1967~\Cite{bunner1967}.\Label{tab:bunner1}}  
\begin{tabular}{cccccc}\hline
 & ~~$p^\prime({\rm N}_2)$~~ & ~~$p^\prime({\rm air})$~ & ~~~~~~~~$\sigma_{no}$~~~~~~~~ 
& ~~~~~~~~$\sigma_{nn}$~~~~~~~~ & ~$\tau_0\times 10^8$~\\ 
 &  \multicolumn{2}{c}{mm Hg} & \multicolumn{2}{c}{cm$^2$} & sec \\ \hline
1N & & & & & \\
$v$=0 & 1.49 & 1.08 & 13 $\times 10^{-15}$ & 4.37 $\times 10^{-15}$ & 6.58 $\pm$ 0.35 \\
2P & & & & & \\
$v = 0$ & 90 & 15 & 2.1 $\times 10^{-15}$ & 1.0 $\times 10^{-16}$ & 4.45 $\pm$ 0.6 \\
~~~~~1 & 24.5 & 6.5 & & 3.5 $\times 10^{-16}$ & 4.93 \\
~~~~~2 & 10.9 & 4.6 & & 8.8 $\times 10^{-16}$ & 4.45 \\
~~~~~3 & 5.4 & 2.5 & & 1.2 $\times 10^{-15}$ & 6.65 \\ 
\hline 
\end{tabular}
\end{table*}

The de-excitation process was subdivided by Bunner into three parts: radiative transition 
to any lower
state, collisional quenching, and internal quenching with the corresponding life times
$\tau_v$, $\tau_c$, and $\tau_i$. In further considerations, Bunner did not
differentiate between the first and last de-excitation channel anymore and introduced the
nomenclature of\footnote{In the
nomenclature of the summary article \Cite{summary},
 the $\tau_0$ from Bunner corresponds to
$\tau_{v^\prime}^r$.}
\begin{equation}
\frac{1}{\tau_0} = \frac{1}{\tau_v}+\frac{1}{\tau_i}.
\end{equation}
The fluorescence efficiency is then defined as $\frac{(\tau_0/\tau_v)}{1+\tau_0/\tau_c}$
photons per excitation. The lifetime due to collisional quenching can be described by
kinetic gas theory: $\tau_c = 1 / (\sqrt{2} N \sigma_{{\rm nn}}\overline{v})$, where
$\sigma_{{\rm nn}}$ is the cross section for nitrogen-nitrogen quenching and
$\overline{v}$ the mean molecular velocity $= \sqrt{(8kT)/(\pi M)}$, with $k$ = Boltzmann
constant, $T$ = temperature of the gas, and $M$ = molecular mass. After several
conversions and introducing $p^\prime$ as a reference pressure, the fluorescence
efficiency is then written as\footnote{Compare with (23) of \Cite{summary}.}
\begin{equation}
{\rm fluorescent~efficiency} = \frac{(\tau_0/\tau_v)}{1 + p/p^\prime}.
\end{equation}
Thus, $\tau_0$ is the fluorescence decay time in the absence of collisional quenching. For
the fluorescence emission in air, Bunner expanded the given definition by a term accounting
for nitrogen-oxygen quenching~\Cite{bunner1967}:
\begin{equation}
\frac{1}{\tau_{no}} = 2 N_o \sigma_{no}\sqrt{\frac{2kT (M_n+M_o)}{\pi M_n M_o}}, 
\end{equation}
which yields a 
\begin{equation}
{\rm fluorescent~efficiency} =
\frac{(\tau_0/\tau_v)}{1+\tau_0(\frac{1}{\tau_{nn}}+\frac{1}{\tau_{no}})}, 
\end{equation}
with $\sigma_{no}$ as the effective nitrogen-oxygen collisional cross section for
de-excitation, $N_o$ is the number of oxygen molecules per cm$^3$, and $M_n$ and $M_o$ are
the nitrogen and oxygen molecular masses.

In \Cite{bunner1967}, Bunner compared several measurements and stressed that the upper
equations are only valid under thin target conditions\footnote{For a definition of thin
target vs.~thick target conditions, please read the accompanying summary
article~\Cite{summary}.}. For high pressure conditions, he 
compared his own measurements performed during the studies for his thesis with those from
Davidson and 
O'Neil~\Cite{davidson_oneil1964} and concludes a fair agreement. As a basis of further
calculations, he summarized deactivation constants for nitrogen and air at a gas
temperature of about 300~K, see Table~\Ref{tab:bunner1}.

Bunner discussed several experiments, e.g.~Gr\"un and
Schopper~\Cite{grun_schopper1954}, Davidson and O'Neil~\Cite{davidson_oneil1964},
Brocklehurst~\Cite{brocklehurst1964} as reviewed in this article in large detail.

In the following, some discussions about quenching and dependences on gaseous conditions
are reviewed. Bunner stressed that the quenching mechanism is not well
understood. The cross sections for collisional quenching seem to be larger than their
molecular radii as given by gas kinetic theory. Furthermore, the cross sections appear to
be generally higher for higher vibrational excitation, except for the 1N system of
nitrogen. It was pointed out that molecular oxygen is a very efficient quencher due to its
permanent dipole moment and the many low-lying energy levels to which excitation energy
may be transferred in a collision~\Cite{bunner1967}. Also some recombination reactions
were discussed, but none of them has important effects on nitrogen fluorescence in the
atmosphere. The 1\% argon content of air might yield additional fluorescence emission
because of a high cross section for argon excitation by electrons. However, experiments on
argon-nitrogen mixtures have verified de-excitations of argon due to quenching
processes. Bunner concluded an expectation of less than 1\% for the argon contribution to
the 2P nitrogen emission in air. The fluorescence yield depends also on temperature and
pressure. Thus, a day-to-day or seasonal variation and a altitude-dependent variation has
to be considered. The deactivation cross sections are taken to be independent of
temperature. The day-to-day variations vary the fluorescence efficiency in the order of
1\%, assuming a mean temperature of 290~K and a 5\% temperature change. Difference of
about 20~K, as between summer and winter, leads to a reduction of the fluorescence yield up
to 4\% for the lower temperature. With increasing altitude, the temperature decreases and
at about 10~km a.s.l.~it is lower by 60~K which reduces the efficiency by 11\%.

\begin{table}[t]
\caption{\Label{tab:bunner2}Predictions for cosmic ray fluorescence in the lower
atmosphere given in 
Bunner, 1967~\Cite{bunner1967}. Here, only the contributions between 300 and 420~nm are
reviewed.}
\begin{minipage}{\linewidth}
\begin{center}
\begin{tabular}{cccccc}\hline
Band~~ & ~~$\lambda$ & ~$E(\lambda_i)$~ & $p^\prime$ & \multicolumn{2}{c}{Sea Level
Efficiency} \\  
 & ~~(\AA) & (\%) & (mm) & ~~~$\times 10^{-4}$\% ~~ & ~Photons/MeV~\\
 \hline
1N & & & & & \\
0-0 & 3914 & .33 & 1.0 & 4.33 & 1.37 \\
2P & & & & & \\
0-0 & 3371 & .0820 & 15 & 15.9 & 4.32 \\
0-1 & 3577 & .0615 & & 11.9 & 3.44 \\
0-2 & 3805 & .0213 & & 4.12 & 1.27 \\
0-3 & 4059 & .0067 & & 1.30 & .48 \\[3pt]
1-0 & 3159 & .050 & 6.5 & 4.3 & 1.09 \\
1-1 & 3339 & .0041 & & .35 & .094 \\
1-2 & 3537 & .0290 & & 2.48 & .71 \\
1-3 & 3756 & .0271 & & 2.31 & .73 \\
1-4 & 3998 & .0160 & & 1.36 & .44 \\[3pt]
2-1 & 3136 & .029 & 4.6 & .96 & .23 \\
2-2 & 3309 & .0020 & & .12 & .032 \\
2-3 & 3500 & .0040 & & .24 & .068 \\
2-4 & 3711 & .0100 & & .60 & .18 \\
2-5 & 3943 & .0064 & & .38 & .12 \\[3pt]
3-2 & 3117 & .005 & 2.5 & .16 & .04 \\
3-3 & 3285 & .0154 & & .50 & .13 \\
3-4 & 3469 & .0063 & & .21 & .06 \\
3-5 & 3672 & .0046 & & .15 & .045 \\
3-6 & 3894 & .003 & & .10 & .03 \\
3-7 & 4141 & .0017 & & .056 & .02 \\ 
\hline 
\end{tabular}
\end{center}
\end{minipage}
\end{table}

Finally, Bunner calculated the fluorescence emission in the lower atmosphere as a
prediction for cosmic ray induced scintillation. The set of best self-consistent quenching
parameters used is reviewed in Tab.~\Ref{tab:bunner1}. Additionally, Bunner applied
absolute fluorescent efficiencies in absence of quenching from Hartman~\Cite{hartman1963},
Bunner~\Cite{bunner1964}, and Davidson and O'Neil~\Cite{davidson_oneil1964}. The agreement
between these measurements is fair. For the calculation, Bunner used a set of weighted
averages for the fluorescence
efficiency for high energy electrons~\Cite{bunner1967}. In
Table~\Ref{tab:bunner2}\footnote{In the nomenclature of the summary article \Cite{summary},
$E(\lambda_i)$ from Bunner corresponds to $\Phi_\lambda^0$, the sea level efficiency in \%
to $\Phi_\lambda$, and the sea level efficiency in Photons/MeV to $Y_\lambda$.},
Bunner's calculations for the fluorescence emission are given which were used by all
cosmic ray experiments as the standard reference for many decades.

\section{R.~O'Neil, G.~Davidson (1968)}
\Label{sec:oneil_davidson}

The former publication of these two authors was embedded in a multi-year project (1964 -
1967) for the 
\emph{Air Force Cambridge Research Laboratories, Office of Aerospace Research, United
States Air Force, Bedford, Massachusetts}. The aim of this project was the interpretation
of the nature of the optical emissions for atmospheric constituents excited by energetic
electrons in a controlled laboratory experiment for understanding ionospheric radiative
phenomena, including natural aurora and airglow emissions as well as the optical emissions
associated with the detonation of a nuclear device. In the final
report~\Cite{oneil_davidson1968}, revised values of the absolute fluorescent efficiencies
published in~\Cite{davidson_oneil1964} were presented. Essentially, the earlier values
were confirmed to the order of 10 to 20 percent. The entire project covers
measurements for thick and thin target in nitrogen and air. 

For the thick target
case\addtocounter{footnote}{-1}\footnotemark[\value{footnote}], fluorescent efficiencies 
were obtained at a pressure of 22~Torr excited by 10~keV 
electrons and at 600~Torr excited by 50~keV electrons. The spectra were observed in the
wavelength region between 2\,000 and 11\,000~\AA{}, but the efficiencies were only
presented above 3\,200~\AA{}, because of impurities in the spectrum from NO $\gamma$
bands between 2\,000 and 3\,200~\AA{}. The band systems observed were the second positive
($C^3\Pi_u \rightarrow B^3\Pi_g$), first positive ($B^3\Pi_g \rightarrow A^3\Sigma_u^+$),
Gaydon-Green and Herman infrared systems of N$_2$ and the first negative system 
($B^2\Sigma_u^+ \rightarrow X^2\Sigma_g^+$) of N$_2^+$. In addition, the NO $\gamma$ 
system ($A^2\Sigma^+ \rightarrow X^2\Pi$), CN violet ($B^2\Sigma^+ \rightarrow
X^2\Sigma^+$), and red systems ($A^2\Pi_i \rightarrow X^2\Sigma^+$) 
were observed as impurities in the target chamber. The revised spectral fluorescent
efficiencies can be found in
Table~\Ref{tab:davidson_oneil1}\addtocounter{footnote}{+1}\footnote{In the nomenclature 
of the summary article \Cite{summary}, $\varepsilon$ from Davidson and O'Neil corresponds to
$\Phi_\lambda$.}. 
 Here, only the
contributions between 320 and 420~nm are repeated.

\begin{table}[t]
\caption{Spectral fluorescent efficiencies $\epsilon$ of nitrogen and air at 600~Torr
given in 
O'Neil and Davidson, 1968~\Cite{oneil_davidson1968}.\Label{tab:davidson_oneil1}} 
\renewcommand{\thefootnote}{\thempfootnote}
\begin{minipage}{\linewidth}
\begin{center}
\begin{tabular}{ccccccc}\hline
 & Molecule~ &~ Band or & \multicolumn{2}{c}{N$_2$ fl.~eff.} &
\multicolumn{2}{c}{Air fl.~eff.} \\
$\lambda$ (\AA{}) & or atom & line(s) & ($\times$10$^{-7}$) & Rem.\footnote{Remarks:
When more 
than one identification is given the fluorescent efficiency refers to the identified
source as a single spectral feature. The symbols indicate the estimated accuracy of the
relative fluorescent efficiencies as follows: $I$, isolated, $\pm$10\%; $SO$, slightly
overlapped, $\pm$20\%; $PO$, partially overlapped, $\pm$35\%. The error in the absolute
measurement is estimated to be an additional $\pm$15\%.} &
  ($\times$10$^{-7}$) & Rem.\footnotemark[\value{mpfootnote}] \\ \hline
3285 & N$_2$ & 2P(3-3) & 44 & $SO$ & 6.4 & $I$ \\
3371 & N$_2$ & 2P(0-0) & 4300 & $I$ & 210 & $I$ \\
3469 & N$_2$ & 2P(3-4) & 11 & $PO$ & 2.6 & $SO$ \\
3501 & N$_2$ & 2P(2-3) & $\ldots$ & $\ldots$ & 2.0 & $PO$ \\
3537 & N$_2$ & 2P(1-2) & 440 & $SO$ & 30 & $SO$ \\
3577 & N$_2$ & 2P(0-1) & 3\,000 & $I$ & 140 & $I$ \\
3672 & N$_2$ & 2P(3-5) & 17 & $PO$ & 1.9 & $SO$ \\
3711 & N$_2$ & 2P(2-4) & 72 & $SO$ & 7.7 & $I$ \\
3756 & N$_2$ & 2P(1-3) & 340 & $I$ & 31 & $I$ \\
3805 & N$_2$ & 2P(0-2) & 1180 & $I$ & 53 & $I$ \\
3853 & CN\footnote{CN violet - main system, transitions (4-4), (3-3), (2-2), (1-1),
(0-0).} & & 200 & $SO$ & $\ldots$ & $\ldots$ \\ 
3914 & N$^+_2$ & 1N(0-0) & 120 & $I$ & 70 & $I$ \\
3943 & N$_2$ & 2P(2-5) & 80 & $PO$ & 4.8 & $PO$ \\
3998 & N$_2$ & 2P(1-4) & 180 & $I$ & 18 & $I$ \\
4059 & N$_2$ & 2P(0-3) & 360 & $I$ & 18 & $I$ \\ 
4142 & N$_2$, CN\footnote{N$_2$ and CN violet - main system, transitions 2P(3-7), CN
(5-6), 
(4-5).} & & 38 & $SO$ & 1.3\footnote{only for the transition 2P(3-7); the CN transitions
have not been observed in air.} & $SO$ \\
4201 & CN, N$_2$\footnote{CN violet - main system and N$_2$, transitions (3-4), (2-3),
2P(2-6).} & & 90 & $SO$ & 3.5\footnote{only for the transition 2P(2-6); the CN transitions
have not been observed in air.} & $I$ \\ \hline 
\end{tabular}
\end{center}
\end{minipage}
\end{table}

The N$_2^+$ first negative system was studied in more detail. The efficiencies were
measured for many combinations of incident electron energy (10 to 60~keV) and nitrogen gas
pressure (22 to 800~Torr). The resulting linear relationship indicates that the
fluorescent efficiency is independent of electron energy and that the Stern-Volmer mechanism
describes the quenching process of the 391.4~nm emission:
\begin{equation}
\frac{1}{\epsilon}=\frac{1}{\epsilon_0}(1+2.2\cdot10^{21}\cdot \sigma \tau P).
\end{equation}
Here, $\epsilon$ is the fluorescent efficiency at any pressure, $\epsilon_0$\footnote{In
the nomenclature of the summary article \Cite{summary}, $\epsilon_0$ from Davidson and O'Neil
corresponds to $\Phi_\lambda^0$.} the
fluorescent efficiency at low pressure where quenching can be neglected, $\sigma$ the
deactivation cross section of the neutral nitrogen molecule in cm$^2$, $\tau$ the lifetime
of the 391.4~nm band in seconds and $P$ the pressure in Torr. From their thin target
measurements, the authors obtained $\epsilon_0 = (6.0 \pm 1.0)\times 10^{-3}$ in nitrogen
for 
excitation by electrons with several hundred eV or more. Applying lifetime measurements
from Bennett and Dalby~\Cite{bennett_dalby1959} with $\tau = (6.58 \pm 0.35)\times
10^{-8}$~s, the N$_2$ collisional deactivation cross section becomes $(5.9 \pm
1.4)\times 10^{-15}$~cm$^2$. Similar measurements were performed in air. The value of
$\epsilon_0$ in air was estimated to be 0.76 the value in N$_2$. Using the result for
the cross section in N$_2$, an O$_2$ quenching cross section of $(1.4\pm 1.0)\times
10^{-14}$~cm$^2$ was obtained. The emission at 391.4~nm was calibrated
absolutely. The relative integrated intensities of the second positive system were
converted into absolute fluorescent efficiencies by a conversion factor based on the known
absolute value for the 391.4~nm band at a given target pressure. For further details
see~\Cite{oneil_davidson1968}. The collisional deactivation cross sections of N$_2$ and
air for the second positive bands can be found in Table~\Ref{tab:davidson_oneil2}.
\begin{table}[t]
\caption{Collisional deactivation cross sections of N$_2$ and
air for the second positive bands of N$_2$ given in
O'Neil and Davidson, 1968~\Cite{oneil_davidson1968}.\Label{tab:davidson_oneil2}} 
\begin{minipage}{\linewidth}
\begin{center}
\begin{tabular}{ccc}\hline
Transition & N$_2$ & Air  \\ 
 &~ $(\times 10^{-16})$~cm$^2$~ &~ $(\times 10^{-15})$~cm$^2$~ \\ \hline  
(0-0) & 0.56 $\pm 0.08$ & 0.95 $\pm$ 0.15 \\ 
(0-1) & 0.50 $\pm 0.08$ & 0.85 $\pm$ 0.13 \\
(0-2) & 0.51 $\pm 0.07$ & 1.0 $\pm$ 0.16 \\
(0-3) & 0.49 $\pm 0.07$ & 1.0 $\pm$ 0.16 \\
(0-4) & 0.50 $\pm 0.08$ & 1.0 $\pm$ 0.16 \\
(0-5) & 0.85 $\pm 0.20$ &  \\ \hline
(1-2) & 1.4 $\pm 0.4$ & 1.5 $\pm$ 0.3 \\
(1-3) & 1.2 $\pm 0.15$ & 1.4 $\pm$ 0.3 \\
(1-4) & 1.4 $\pm 0.2$ & 1.3 $\pm$ 0.2 \\
(1-6) & 1.3 $\pm 0.2$ & 1.6 $\pm$ 0.6 \\
(1-7) &  & 1.3 $\pm$ 0.2 \\ \hline
(2-4) & 1.5 $\pm 0.4$ & 1.1 $\pm$ 0.2 \\
(2-6) &  & 1.5 $\pm$ 0.3 \\
(2-7) &  & 1.1 $\pm$ 0.2 \\
 \hline 
\end{tabular}
\end{center}
\end{minipage}
\end{table}
Analyzing several Stern-Volmer plots of the second positive bands in air and N$_2$, values
for the fluorescent efficiencies $\epsilon_0$ at pressures low enough to exclude quenching
were derived by O'Neil and Davidson. The results for the first four band systems of
the second positive system are repeated in Table~\Ref{tab:davidson_oneil3}.
\begin{table}[t]
\caption{Fluorescent efficiencies $\epsilon_0$ for the N$_2$ second positive bands excited
by energetic electron incident on N$_2$ and air given in
O'Neil and Davidson, 1968~\Cite{oneil_davidson1968}.\Label{tab:davidson_oneil3}} 
\renewcommand{\thefootnote}{\thempfootnote}
\begin{minipage}{\linewidth}
\begin{center}
\begin{tabular}{cccc}\hline
 Transition & \multicolumn{2}{c}{$\epsilon_0 (\times 10^{-5})\footnote{The fluorescent
efficiencies are based on measurements in N$_2$ and air extrapolated to low pressure where
collisional deactivation is an insignificant depopulating process. The probable error does
not include the error in the absolute measurement which is estimated to be an additional
$\pm$ 15\%. }$} & Ratio \\
 &~~~~~~ N$_2$~~~~~~ &~~~~~~ Air~~~~~~ & $\epsilon_0({\rm N}_2)/\epsilon_0(Air)$~ \\
\hline 
0-0 & 186 $\pm$ 16 & 112 $\pm$ 15 & 1.7 $\pm$ 0.3 \\
0-1 & 126 $\pm$ 10 & 82 $\pm$ 11 & 1.5 $\pm$ 0.2 \\
0-2 & 48 $\pm$ 4 & 31 $\pm$ 4 & 1.5 $\pm$ 0.2 \\
0-3 & 14.6 $\pm$ 1.2 & 9.9 $\pm$ 1.5 & 1.5 $\pm$ 0.3 \\
0-4 & 4.0 $\pm$ 0.3 & 2.9 $\pm$ 0.4 & 1.5 $\pm$ 0.3 \\
0-5 & 1.1 $\pm$ 0.2 & & \\ \hline
1-2 & 38 $\pm$ 8 & 29 $\pm$ 6 & 1.3 $\pm$ 0.4 \\
1-3 & 26 $\pm$ 2 & 24 $\pm$ 4 & 1.1 $\pm$ 0.2 \\
1-4 & 18 $\pm$ 2 & 13 $\pm$ 2 & 1.4 $\pm$ 0.3 \\
1-6 & 1.54 $\pm$ 0.14 & 2.0 $\pm$ 0.7 & 0.8 $\pm$ 0.3 \\
1-7 & & 0.37 $\pm$ 0.07 &  \\ \hline
2-4 & 8.4 $\pm$ 1.6 & 5.1 $\pm$ 0.9 & 1.6 $\pm$ 0.4 \\
2-6 & & 3.1 $\pm$0.7 & \\
2-7 & & 0.81 $\pm$ 0.13 & \\
2-8 & 0.40 $\pm$ 0.08 & 0.35 $\pm$ 0.11 & 1.1 $\pm$ 0.4 \\ \hline
3-3 & 5.6 $\pm$ 1.2 & 5.1 $\pm$ 0.7 & 1.1 $\pm$ 0.4 \\
3-4 & 1.0 $\pm$ 0.3 & 1.1 $\pm$ 0.3 & 0.9 $\pm$ 0.4 \\
3-5 & 2.0 $\pm$ 0.6 & 1.3 $\pm$ 0.4 & 1.5 $\pm$ 0.6 \\
3-7 & & 1.1 $\pm$ 0.4 & \\
3-8 & 0.68 $\pm$ 0.12 & 0.52 $\pm$ 0.17 & 1.3 $\pm$ 0.5 \\ \hline 
\end{tabular}
\end{center}
\end{minipage}
\end{table}
For the thin target case, the gas pressure was reduced so that the incoming electrons
lost only a small fraction of their energy in crossing the observation region of the target
chamber. A distinction between primary and secondary electron excitation was based on the
spatial distribution of the emittance. However, detailed studies have been performed
mainly for the Meinel bands of N$^+_2$~\Cite{oneil_davidson1968}.

\section{Cross section and lifetime data between 1969 and 2000}
\Label{sec:cross_section1969_2000}

After these fundamental experiments on fluorescence emission for cosmic ray experiments,
Sect.~\Ref{sec:grun_schopper} to \Ref{sec:oneil_davidson},
several detailed studies on sets of cross sections and radiative lifetimes have been
published by many authors. Here only those of them are reviewed who are cited in the
accompanying article~\Cite{summary} in
discussions about fluorescence emission of extensive air showers.

\subsection{Stanton and St.~John, (1969)}

Stanton and St.~John measured in 1969 the optical cross sections for, amongst others, the
first negative bands of N$_2^+$ and introduced an apparent cross section for
excitations~\Cite{stanton_stjohn1969}. These apparent cross sections $Q^\prime(v^\prime)$
are defined as the sum of 
all optical cross sections of all bands which have the same upper vibrational state
$v^\prime$. For low pressures and beam currents, where the optical cross sections are
independent of these variables, $Q^\prime(v^\prime) =
\sum_v^{\prime\prime}Q(v^\prime,v^{\prime\prime})$. $Q^\prime(v^\prime)$ also is
equal to the true cross section $Q(v^\prime)$ of the excitation of $v^\prime$ plus the
sum of the optical cross sections of all bands cascading to $v^\prime$ from higher
energy states. The excitation functions were measured for the first negative system
of N$_2^+$ up to energies of 450~eV and the functions show broad maxima at about 120~eV
for the (0,0) and (1,0) bands. Frank-Condon factors indicated that the measured optical
cross sections account for over 99\% of $Q^\prime(0)$ and $Q^\prime(1)$. The maximum
apparent cross sections are given in Table~\Ref{tab:stanton_stjohn}.
\begin{table}[t]
\caption{Maximum apparent cross sections of the first 4 vibration states of the N$_2^+
B^2\Sigma_u^+$ electronic state as given
in~\Cite{stanton_stjohn1969}.\Label{tab:stanton_stjohn}}  
\begin{minipage}{\linewidth}
\begin{center}
\begin{tabular}{cc}\hline
$~v$~~~ & $Q^\prime(v)(10^{-18}$~cm$^2$) \\ \hline
0 & 22.3 \\
1 & 2.6 \\
2 & 0.21 \\
3 & 0.09 \\ \hline
\end{tabular}
\end{center}
\end{minipage}
\end{table}

\subsection{Dotchin et al., (1973)}

In 1973, Dotchin et al.~\Cite{dotchin_etal1973} presented radiative lifetime measurements
using a pulsed-proton beam and target gas pressures between 1 and 200 mTorr. They studied
two gases, nitrogen and carbon monoxide. For the first negative system of N$^+_2$, the
measurements were performed at the 3914~\AA{}~(0,0) and the 4278~\AA{}~(0,1) bands. The
mean zero-pressure intercept yielded (60.4 $\pm$0.4)~nsec. The lifetimes for the second
positive system of nitrogen states were investigated using the 3805~\AA{}~(0,2),
4059~\AA{}~(0,3), 3755~\AA{}~(1,3), 3998~\AA{}~(1,4), and the 3711~\AA{}~(2,4)
transitions. The mean lifetime for $v^\prime = 0$ was determined to be (40.4 $\pm$
0.5)~nsec, for $v^\prime = 1$ to (40.6 $\pm$0.5)~nsec, and for $v^\prime = 2$ to 
(38.5 $\pm$0.6)~nsec. 

\subsection{Lillicrap, (1973)}

In the same year, a NASA report about the temperature dependence of collisional quenching
was presented~\Cite{lillicrap1973}. Lillicrap investigated the 391.4~nm line of nitrogen
and the 501.6~nm line 
of helium. The work was performed because of interest in hypersonic flight at high
altitudes and here the high-speed, low-density flows were to be studied. The quenching
behavior of nitrogen was measured in the temperature interval from 78~K to 300~K. In
their analysis, only collisions between excited molecules and ground-state molecules of
the same species were assumed. The gas was excited by a 30~keV electron beam of
200~$\mu$A. The cooling was realized by a heat exchanger filled with 2-methylbutane for the
temperatures between 300~K and 113~K and with liquid nitrogen to obtain temperatures close
to 78~K. For analyzing the quenching cross sections, the densities were limited
between 1$\times 10^{22}$ and $12 \times 10^{22}$ molecules/m$^3$, because of not
conceived effects at lower densities. The given uncertainties are only those of
individual measurements and no reliable standard deviations. The measured cross sections
are given in Table~\Ref{tab:lillicrap}.
\begin{table}[t]
\caption{Cross sections for quenching of the $v^\prime$ = 0 level of the N$^+_2$
B$^2\Sigma^+_u$ state of N$_2$ as given in~\Cite{lillicrap1973}.\Label{tab:lillicrap}}  
\begin{minipage}{\linewidth}
\begin{center}
\begin{tabular}{cc}\hline
$T$ (K)~~~ & $\sigma_{\rm{NN}}$ (m$^2$) \\ \hline 
296 & 4.8 $\pm$ 0.2 $\times 10^{-19}$ \\
247 & 5.6 $\pm$ 0.3 $\times 10^{-19}$ \\
208 & 6.7 $\pm$ 0.3 $\times 10^{-19}$ \\
168 & 7.7 $\pm$ 0.2 $\times 10^{-19}$ \\
162 & 8.7 $\pm$ 0.6 $\times 10^{-19}$ \\
123 & 9.8 $\pm$ 0.7 $\times 10^{-19}$ \\
118 & 10.0 $\pm$ 0.8 $\times 10^{-19}$ \\
78 & 11.0 $\pm$ 0.9 $\times 10^{-19}$ \\ \hline
\end{tabular}
\end{center}
\end{minipage}
\end{table}

\subsection{Lofthus and Krupenie, (1977)}

For a really comprehensive review about ``The Spectrum of Molecular Nitrogen'', the reader
is referred to Lofthus and Krupenie, 1977~\Cite{lofthus_krupenie1977}. Within this review
and 
compilation, data are presented of the observed and predicted spectroscopic data on the
molecule N$_2$ and its ions N$_2^-$, N$_2^+$, N$_2^{2+}$, and the molecule N$_3$. Amongst
others, radiative lifetimes and Franck-Condon integrals are given. For example the radiative
lifetimes as weighted averages of many experiments are given for the second positive 0-0
transition with $(3.66\pm 0.05)\cdot 10^{-8}$~sec and for the first negative 0-0 band with
$(6.25\pm 0.08)\cdot 10^{-8}$~sec.

\subsection{Itikawa, (1986)}

In 1986, Itikawa et al.~\Cite{itikawa_etal1986} published again a compilation of data
about nitrogen with strong emphasis on cross sections for collisions of electrons and
photons. For electron collisions, the processes of total scattering, elastic scattering,
momentum transfer, excitations of rotational, vibrational and electronic states,
dissociation, and ionization are considered and presented graphically. Spectroscopic and
other properties of the nitrogen molecule are summarized. Also in the work by Itikawa et
al.~\Cite{itikawa_etal1986}, radiative lifetimes are presented, however these numbers have
been calculated theoretically. In Table~\Ref{tab:itikawa}, the values for the second
positive and first negative system of nitrogen are reviewed.
\begin{table}[t]
\caption{Lifetimes (in units of 10$^{-8}$ seconds) of electronic states of N$_2$ and
N$_2^+$ as given in~\Cite{itikawa_etal1986}.\Label{tab:itikawa}}  
\begin{minipage}{\linewidth}
\begin{center}
\begin{tabular}{ccc}\hline
$~v^\prime$~~~ & C$^3\Pi_u \rightarrow$B$^3\Pi_g$~ &~ B$^2\Sigma_u^+ \rightarrow$
X$^2\Sigma_g^+$ \\ \hline 
0 & 3.67 & 5.52 \\
1 & 3.65 & 5.38 \\
2 & 3.69 & 5.30 \\
3 & 3.77 & 5.27 \\ 
4 & 3.94 & 5.27 \\
5 & & 5.33 \\ \hline
\end{tabular}
\end{center}
\end{minipage}
\end{table}

\subsection{Gilmore et al., (1992)}

A more theoretical analysis was published by Gilmore et al.~in
1992~\Cite{gilmore}. In this comprehensive report, calculations on the
Franck-Condon factors, $r$-centroids (with $r$ being the internuclear distance),
electronic transition moments, and Einstein 
coefficients for many nitrogen and oxygen band systems are presented. The Einstein
coefficients or radiative transition probabilities can be used for calculating
emission spectra produced by electrons incident in air. The Franck-Condon factors are the
vibrational overlap integrals and are used for calculating the branching ratios for
populating various vibrational levels when an electronic state is excited from the ground
state by electron impact~\Cite{gilmore}. The radiative transition parameters of
38 band systems considered in that report are listed in several tables. Also the radiative
lifetimes of 14 N$_2$, N$_2^+$, and O$_2^+$ states are calculated and listed as a function
of vibration level. Since this comprehensive report comprises of 86 pages of tables, no
data are reviewed here. 

\subsection{Fons et al., (1996)}

The idea of the apparent excitation cross sections was seized upon again by Fons et al.~in
1996~\Cite{fons_etal1996}. The absolute optical emission cross sections were measured
for the second positive system of nitrogen for many transitions from $v^\prime$ =
0,1,2,3,4 and $v^\prime$ up to 9. The electron beam energy was modulated from threshold
up to 600~eV and the gas pressure was fixed at about 4~mTorr. The estimated uncertainty
for each cross section is also given in~\Cite{fons_etal1996}. For the (0,0) band it is
13\%, but for many other bands, the uncertainty is roughly 20\%. The apparent electron
excitation cross section of the $C^3\Pi_u(v^\prime)$ vibrational level is given as the
sum of the optical emission cross sections of  
the ($v^\prime ,v^{\prime\prime}$) bands over $v^{\prime\prime}$, because the
$C\rightarrow B$ transition is the only radiative decay channel of the $C^3\Pi_u$
state. The absolute and relative values are reviewed in Table~\Ref{tab:fons} and compared
with relative Franck-Condon factors calculated by Gilmore et al.~\Cite{gilmore}.
\begin{table}[t]
\caption{Apparent excitation cross sections (in units of 10$^{-18}$~cm$^2$) for the
$C^3\Pi_u(v^\prime)$ vibrational levels, $Q_{app}(Cv^\prime)$, at incident energies
corresponding to the peak of the excitation functions as given
in~\Cite{fons_etal1996}. The relative values of these cross
sections in the third column are compared to the relative Franck-Condon (FC)
factors.\Label{tab:fons}}   
\begin{minipage}{\linewidth}
\begin{center}
\begin{tabular}{cccc}\hline
 & & Relative & Relative \\
$~v^\prime$~ & ~$Q_{app}(Cv^\prime)$~ & ~$Q_{app}(Cv^\prime)$~ & FC factors  \\ \hline 
0 & 21.7 & 1.00 & 1.00 \\ 
1 & 15.2 & 0.70 & 0.57 \\
2 & 5.57 & 0.26 & 0.19 \\
3 & 1.76 & 0.081 & 0.055 \\ 
4 & 0.88 & 0.041 & 0.014\\ \hline
\end{tabular}
\end{center}
\end{minipage}
\end{table}
The comparison between the relative $Q_{app}(Cv^\prime)$ values and the relative
Franck-Condon factors give rise to the discussion about the validity of the Franck-Condon
approximation. If the cascade from higher levels is assumed to be small in comparison with
the direct excitation cross section, the apparent excitation cross section can be compared
with the appropriate Franck-Condon factors. For $v^\prime$ = 0,1,2 the numbers agree
rather well, but for $v^\prime$ = 2 to 3 an abrupt decrease both in the relative
Franck-Condon factor (or the direct excitation cross section) and in the relative apparent
cross section can be seen. The authors of \Cite{fons_etal1996} estimated the contribution
from the missing optical emission cross section to the apparent cross sections to be no
more than 1\% for $v^\prime$=0,1,2, no more than 4.5\% for $v^\prime$=3, and
approximately 10\% for $v^\prime$=4. Finally, they concluded that the deviation of the
apparent excitation cross sections from the Franck-Condon relation is consistent with the
cascade description and that it does not necessarily signify a breakdown of the
Franck-Condon picture~\Cite{fons_etal1996}.

\subsection{Pancheshnyi et al., (1997-2000)}

Between 1997 and 2000 Pancheshnyi et al. published four reports about rate constants and
lifetimes of nitrogen~\Cite{pancheshnyi_etal}. Here, only the latest one (2000) is
reviewed since 
former data are updated therein. The
measurements were performed using emission spectroscopy with sub-nanosecond temporal
resolution. The spectral range of the optical system was between 250 and 600~nm. Gas
pressures ranged from 0.05 to 30~Torr at a temperature of 295 $\pm$ 5~K. The gas was
either pure nitrogen or nitrogen mixtures with oxygen, hydrogen or oxygen and hydrogen. It
was measured that the excitation occurs up to 25~ns at 1~Torr pressure
after discharge and up to 5-6~ns at high pressures. The values of the radiative lifetimes
$\tau$ 
were obtained by extrapolation to zero pressure and the quenching rates were determined
from the slope of the straight line 1/$\tau$ vs.~pressure~\Cite{pancheshnyi_etal},
(2000). The uncertainties of the given values are given with 5-15\% on average. In
Table~\Ref{tab:pancheshnyi}, the radiative lifetimes and the deactivation rate constants
are summarized.
\begin{table}[t]
\caption{Radiative lifetimes $\tau_0$ (in 10$^{-9}$~s) and rate constants $k_q^M$ of
deactivation by molecule $M$ (in 10$^{-10}$~cm$^3$s$^{-1}$) for N$_2(C^3\Pi_u,v=0\ldots
4)$  as given in~\Cite{pancheshnyi_etal}, (2000).\Label{tab:pancheshnyi}}    
\begin{minipage}{\linewidth}
\begin{center}
\begin{tabular}{ccccc}\hline
~$v$~ & ~~~~~$\tau_0$~~~~~ & ~~~~~~$k_q^{{\rm N}_2}$~~~~~~  & ~~~~~~$k_q^{{\rm
O}_2}$~~~~~~  & ~~~~~~$k_q^{{\rm H}_2{\rm O}}$~~ \\ 
\hline 
0 & 42 $\pm$ 2 & 0.13 $\pm$ 0.02 & 3.0 $\pm$ 0.3 & 3.9 $\pm$ 0.4 \\ 
1 & 41 $\pm$ 3 & 0.29 $\pm$ 0.03 & 3.1 $\pm$ 0.3 & 3.7 $\pm$ 0.4 \\
2 & 39 $\pm$ 4 & 0.46 $\pm$ 0.06 & 3.7 $\pm$ 0.5 & 4.0 $\pm$ 0.6 \\
3 & 41 $\pm$ 5 & 0.43 $\pm$ 0.06 & 4.3 $\pm$ 0.6 & 4.5 $\pm$ 0.7 \\ \hline 
\end{tabular}
\end{center}
\end{minipage}
\end{table}

\section{Fluorescence yield measurements between 1970 and 1996}
\Label{sec:yield_1970_1996}

During mainly the same period as covered in Sec.~\Ref{sec:cross_section1969_2000}, only
three well-known measurements on nitrogen fluorescence yield were performed.

\subsection{Hirsh et al., (1970)}

The first publication was by Hirsh et al.~in 1970~\Cite{hirsh1970}. The absolute fluorescence
efficiency of the first negative (0,0) transition of N$_2^+$ was measured in nitrogen
and in an air-like N$_2$:O$_2$ mixture. The incident electron beam covered an energy range
from 0.65 to 1.6~MeV, which had not been studied before then. Under the experimental conditions
described in \Cite{hirsh1970}, the incident primary and all resulting secondary electrons
ionize the ground-state N$_2$ molecule. Approximately one 391.4~nm photon is produced
for every 15 electron-ion 
pairs. For the primary electron it was a thin target
condition, but for the slow secondary electrons it was a thick target. 10\% of the ionized
molecules are formed in the 
$B^2\Sigma_u^+(v^\prime=0)$ state which de-excites by the emission of 391.4~nm
photons.  The resultant fluorescence efficiency\footnote{In the nomenclature of the
summary article \Cite{summary}, this parameter corresponds to $\Phi_\lambda^0$.}, defined
as the power radiated by the gas in 
391.4~nm photons per unit power deposited in the gas by the electron beam, was
determined to (4.75 $\pm$ 1.5)$\times 10^{-3}$ in air and to (6.0 $\pm$ 1.59)$\times
10^{-3}$ in nitrogen. A scan in energy indicated that the fluorescence 
efficiency is energy-independent from near threshold energy to 1.65~MeV. A pressure scan
followed by an extrapolation to zero pressure yielded a reciprocal lifetime of (1.52 $\pm$
0.08) $\times 10^7$~sec$^{-1}$ for the $B^2\Sigma$ state. The cross sections for quenching
were found to be (65 $\pm$ 4) $\times 10^{-16}$~cm$^2$ for nitrogen-nitrogen collisions and
(109 $\pm$ 45) $\times 10^{-16}$~cm$^2$ for nitrogen-oxygen collisions.

\subsection{Mitchell, (1970)}

In the same year, Mitchell published fluorescence efficiencies and collisional
deactivation rates for the second positive bands of N$_2$ and the first negative band of
N$^+_2$~\Cite{mitchell1970}. He used soft monochromatic X-rays from the energy range
between 0.9 and 8.0~keV. All measurements were performed either in pure nitrogen or in an
air-like mixture of 20\% O$_2$ and 80\% N$_2$ at pressures ranging from 
0.30~Torr to 600~Torr. Describing the data by the Stern-Volmer relationship
\begin{equation}
1/\eta =(1/\eta_0)(1+KP),
\end{equation}
where $P$ is the pressure in Torr and $K$ is the quenching constant in Torr$^{-1}$, the 
fluorescence efficiencies\footnotemark[\value{footnote}] were found as given in
Table~\Ref{tab:mitchell}. 
\begin{table}[t]
\caption{Fluorescence efficiencies as given in~\Cite{mitchell1970}.\Label{tab:mitchell}}    
\begin{minipage}{\linewidth}
\begin{center}
\begin{tabular}{cccc}\hline
~wavelength~ & ~~~~~~~Transition~~~~~ & \multicolumn{2}{c}{Efficiencies} \\
 (in \AA{}) & & ~~~Nitrogen~~~  & ~~~Air~~ \\ 
\hline 
3914 & N$_2^+$ 1st neg.~(0,0) & 0.66\% & 0.53\% \\ 
3371 & N$_2$ 2nd pos.~(0,0) & 0.26\% & 0.21\% \\
3805 & N$_2$ 2nd pos.~(0,2) & 0.062\% & 0.048\% \\
4060 & N$_2$ 2nd pos.~(0,3) & 0.018\% & 0.014\% \\ \hline 
\end{tabular}
\end{center}
\end{minipage}
\end{table}
A variation of energy in the given range confirmed that the efficiencies are
independent of the X-ray energy. From theory it was known that the deactivation rates
should be the same 
for all bands with the same upper vibrational state. Mitchell could show experimentally
that the $K$ values for the second positive $v^\prime=0$ bands are the same. Thus he
derived a rate constant for the first negative ($v^\prime$=0) system for
nitrogen-nitrogen quenching of 4.53 $\times 10^{-10}$~cm$^3$/sec and for nitrogen-oxygen
quenching of 7.36 $\times 10^{-10}$~cm$^3$/sec. For the second positive ($v^\prime$=0)
system, Mitchell obtained for nitrogen-nitrogen quenching a rate constant of 1.12 $\times
10^{-10}$~cm$^3$/sec and for nitrogen-oxygen quenching of 3.12 $\times
10^{-10}$~cm$^3$/sec~\Cite{mitchell1970}. These numbers were calculated applying the
lifetime $\tau_0$ = 6.58 $\times 10^{-8}$~sec from Bennett and
Dalby~\Cite{bennett_dalby1959}. 

\subsection{Kakimoto et al., (1996)}

Finally in 1996, the aspects of nitrogen fluorescence emission were again discussed in the
context of extensive air shower detection by fluorescence telescopes. Kakimoto et
al.~studied the fluorescence yield\footnote{In the nomenclature of the
summary article \Cite{summary}, this parameter corresponds to $\epsilon_\lambda$.} with
electron energies between 1.4 and 1000~MeV~\Cite{kakimoto1996}. 
The altitude dependence in the Earth's atmosphere was parameterized and the proportionality
between yield and energy deposit of incident particles was checked. The experiment was
split into two parts. The first part comprised of a 1~mCi $^{90}$Sr source which provided
1.4~MeV electrons for exciting either N$_2$ gas or dry air. In the second part, the
exciting electrons were extracted from a synchrotron with energies of 300, 650, and
1000~MeV. The authors plotted the fluorescence yield vs.~electron energy into one plot
together with the Bethe-Bloch curve for energy deposit by electrons. The absolute scale
was adjusted at 1.4~MeV. The agreement between the curve and the data points is fair, thus
Kakimoto et al.~concluded the proportionality between fluorescence yield 
and energy deposit. The fluorescence efficiency\footnote{In the nomenclature of the
summary article \Cite{summary}, this parameter corresponds to $\Phi_\lambda$.} for 1.4~MeV
electrons at 
600~mm~Hg in air was 
measured at three wavelengths. Defined as the radiated energy
divided by the energy loss in the observed medium, the efficiency was given at 337~nm as
2.1 $\times 10^{-5}$, at 
357~nm as 2.2 $\times 10^{-5}$, and at 391~nm as 0.84 $\times
10^{-5}$~\Cite{kakimoto1996}. For applying their experimental results to extensive air
shower reconstruction, Kakimoto et al.~parameterized the fluorescence yield in dependence
of energy and altitude as
\begin{equation}
\begin{split}\Label{eq:kakimoto}
{\rm yield} &= \frac{(dE/dX)}{(dE/dX)_{1.4~MeV}}\\ 
&\times \rho\biggl\{\frac{A_1}{1+\rho
B_1\sqrt{T}}+\frac{A_2}{1+\rho B_2\sqrt{T}}\biggr\} ,
\end{split}
\end{equation} 
where $dE/dX$ is the electron energy loss, density $\rho$ in kg/m$^3$, temperature $T$ in
Kelvin, and 
$(dE/dX)_{1.4~MeV}$ is the $dE/dX$ evaluated at 1.4~MeV. The experimentally derived
constants are reviewed in Table~\Ref{tab:kakimoto}.
\begin{table}[t]
\caption{Constants $A_1,~A_2,~B_1~{\rm and}~B_2$ used in (\Ref{eq:kakimoto}) as given
in~\Cite{kakimoto1996}.\Label{tab:kakimoto}}     
\begin{minipage}{\linewidth}
\begin{center}
\begin{tabular}{cc}\hline
~$A_1$~~~~ & 89.0 $\pm$ 1.7~m$^2$kg$^{-1}$ ~~ \\
~$A_2$~~~~ & 55.0 $\pm$ 2.2~m$^2$kg$^{-1}$ ~~ \\
~$B_1$~~~~ & 1.85 $\pm$ 0.04~m$^3$kg$^{-1}$K$^{-1/2}$ ~~ \\
~$B_2$~~~~ & 6.50 $\pm$ 0.33~m$^3$kg$^{-1}$K$^{-1/2}$ ~~ \\
 \hline 
\end{tabular}
\end{center}
\end{minipage}
\end{table}

\section*{Acknowledgments}

The authors acknowledge the support of
the Spanish Ministry of Science and Education MEC (FPA2006-12184-C02-01), Comunidad de
Madrid 
(Ref.: 910600) and CONSOLIDER program, and of the German Research Foundation (DFG) (KE
1151/1-1 and KE 1151/1-2). Further support has been granted by the Radboud 
Universiteit Nijmegen as well as by the Universit\"at Karlsruhe (TH) and the
Forschungszentrum Karlsruhe GmbH which are currently merging their activities in the
Karls\-ruhe Institute of Technology (KIT). The authors would like to thank C.~Escobar,
C.~Field, M.~Fraga, K.~Martens, M.~Nagano, J.~Ridky, J.~Rosado, and A.~Ulrich for valuable
comments on the manuscript.


\end{document}